# Investigating the fidelity of explainable artificial intelligence methods for applications of convolutional neural networks in geoscience

by

Antonios Mamalakis[1*], Elizabeth A. Barnes[1], Imme Ebert-Uphoff [2,3]

[1] Department of Atmospheric Science, Colorado State University, Fort Collins, CO
[2] Department of Electrical and Computer Engineering, Colorado State University, Fort Collins, CO
[3] Cooperative Institute for Research in the Atmosphere, Colorado State University, Fort Collins, CO

*email: amamalak@rams.colostate.edu

**Abstract**

Convolutional neural networks (CNNs) have recently attracted great attention in geoscience due to their ability to capture non-linear system behavior and extract predictive spatiotemporal patterns. Given their black-box nature however, and the importance of prediction explainability, methods of explainable artificial intelligence (XAI) are gaining popularity as a means to explain the CNN decision-making strategy. Here, we establish an intercomparison of some of the most popular XAI methods and investigate their fidelity in explaining CNN decisions for geoscientific applications. Our goal is to raise awareness of the theoretical limitations of these methods and gain insight into the relative strengths and weaknesses to help guide best practices. The considered XAI methods are first applied to an idealized attribution benchmark, where the ground truth of explanation of the network is known *a priori*, to help objectively assess their performance. Secondly, we apply XAI to a climate-related prediction setting, namely to explain a CNN that is trained to predict the number of atmospheric rivers in daily snapshots of climate simulations. Our results highlight several important issues of XAI methods (e.g., gradient shattering, inability to distinguish the sign of attribution, ignorance to zero input) that have previously been overlooked in our field and, if not considered cautiously, may lead to a distorted picture of the CNN decision-making strategy. We envision that our analysis will motivate further investigation into XAI fidelity and will help towards a cautious implementation of XAI in geoscience, which can lead to further exploitation of CNNs and deep learning for prediction problems.

## 1. Introduction

In recent years, convolutional neural networks (CNNs) and deep learning in general have seen great application in a plethora of problems in geoscience (Lary et al., 2016; Karpatne et al., 2018; Reichstein et al., 2019), ranging from solid earth science (Bergen et al., 2019), marine science and hydrology (Shen, 2018; Sit et al., 2020) to climate science and meteorology (Barnes et al., 2019; Rolnick et al., 2019; Ham et al., 2019). The popularity of CNNs has risen mainly due to their ability to capture non-linear system behavior and to extract predictive spatiotemporal patterns (LeCun et al., 2015), which makes them of particular interest to geoscientists. Another important reason is the increasing availability of observational and simulated data in this decade (Overpeck et al., 2011; Guo, 2017; Agapiou, 2017; Reinsel et al., 2018) that helps meet the requirement to train CNNs with large datasets.

Despite their potential, an important issue regarding the application of CNNs in the geosciences is their black-box nature, which makes it hard for scientists to interpret predictions and assess the model from a physical perspective, i.e., beyond using prediction performance as the only criterion. The interpretability issue is considered a key issue for deep learning in general, and it has prompted the emergence of a new subfield in computer science, namely eXplainable Artificial Intelligence (XAI; Buhrmester et al., 2019; Tjoa and Guan, 2019; Das and Rad, 2020). The goal of XAI methods is to explain, in a post-prediction setting (typically referred to as *post-hoc explanation*), the decision strategy of a model that otherwise is inherently not interpretable. One common way to do this is to highlight the most important variables in the input space (typically referred to as *features*) that helped the model make a specific prediction. These methods are referred to as "local" XAI methods because they focus on a specific prediction, in contrast to "global" XAI methods that identify important features across all samples (Buhrmester et al., 2019).

XAI methods have already proven to be of great utility for explaining black-box models in computer science and beyond, and they have seen recent application in geoscience too (McGovern et al., 2019; Ebert-Uphoff and Hilburn, 2020; Toms et al., 2020; Mamalakis et al., 2022). Specifically, recent work shows how XAI can help calibrate model trust (Sonnewald and Lguensat, 2021; Mayer and Barnes, 2021; Hilburn et al., 2021; Keys et al., 2021), identify ways to fine-tune models that are performing poorly (Ebert-Uphoff and Hilburn, 2020), and also accelerate learning new science (Barnes et al., 2020; Toms et al., 2021). The results of these recent studies indicate that XAI can be a real game-changer for prediction and classification problems (Mamalakis et al., 2022) and can help further exploit the potential of deep learning in geoscience in our era of big data.

Despite the above, many XAI methods have been shown to exhibit theoretical and practical limitations in explaining black-box models (Sundararajan et al., 2017; Kindermans et al., 2017b; Ancona et al., 2018; Rudin, 2019; Dombrowski et al., 2020; Zhou et al., 2022). Moreover, XAI results are not typically assessed on the basis of a ground truth of explanation, but rather, are based on the subjective evaluation by the analyst/scientist about whether or not the explanation is physically reasonable. However, even if an explanation makes physical sense to a human, it does not necessarily mean that this is the strategy the model in question is actually using (and vice versa). In other words, the human perception of an explanation alone is not a solid criterion for assessing its trustworthiness. Also, what physically makes sense depends on the *a priori* understanding of the problem that the scientist has, and thus, might be different across individuals, especially in problem settings of high levels of complexity. The theoretical and practical limitations of XAI methods, together with the issue of subjectivity in their assessment that may propagate individual biases, have been recognized in the literature (Leavitt and Morcos, 2020) and call for a more objective and systematic investigation of XAI methods' fidelity for a range of different applications and model architectures.

In an effort to introduce more objectivity in the assessment of XAI methods for geosciences, our group proposed a generic approach to develop simple *attribution benchmark datasets* for benchmarking XAI methods (Mamalakis et al., 2021). Attribution benchmark datasets consist of synthetic inputs and outputs, where the functional relationship between the two is known. This allows for deriving the ground truth of what the explanation of the network should look like for each prediction. In this way, the assessment of XAI methods is no longer based on subjective criteria, but rather it is based on the direct comparison of the XAI results to the ground truth of the explanation. As a first example, Mamalakis et al. (2021) generated a large attribution benchmark inspired from a climate prediction setting and applied XAI methods to explain the predictions of a fully-connected neural network. Other studies have also developed similar benchmarks in the field of computer science (Arras et al., 2021; Zhou et al., 2022).

Here, we build on previous studies that deal with the assessment and the benchmarking of XAI methods and we shift our focus to *convolutional* neural networks, with the aim to investigate XAI fidelity in CNN applications in the geosciences. Our goal is to raise awareness of the theoretical limitations of XAI methods and gain insight into the relative strengths and weaknesses to help guide best practices. We focus on some of the most popular XAI methods (e.g., Gradient, Smooth Gradient, Integrated Gradients, Layer-wise Relevance Propagation, among many others) and apply them to explain the predictions of CNNs for two specific classification problems. First, we consider an idealized attribution benchmark dataset, where the CNN is trained to classify pictures of circular and square frames depending on which class of frames covers more area. The simplicity of the prediction task allows us to derive the ground truth of the explanation and assess XAI methods in an objective manner. Thus, this first problem helps us gain insight into limitations that might be overlooked in cases where no ground truth of the explanation is available. In the second problem, we consider a prediction setting with a climate-related task, namely, predicting the number of the atmospheric rivers in daily snapshots of climate simulations. In this setting there is no ground truth of the explanation, as is the case in most geophysical studies. The second problem aims to validate our insights about XAI in a more climate-related setting and illustrate how explanations should be regarded and interpreted in order to avoid reaching false conclusions about the strategy of the network.

In section 2, we introduce the two datasets, discuss CNN architectures and prediction performance, and describe the XAI methods considered in the study. In section 3, we present and discuss the results of the XAI methods when applied to explain the CNN decision strategy, and in section 4, we state our conclusions.

## 2. Data and Methods

### 2.1. Synthetic Attribution Benchmark

For our first classification problem we develop and use a synthetic attribution benchmark dataset to objectively assess XAI methods. An attribution benchmark consists of a synthetic input **X** and a synthetic output Y, with the latter being a known function $F$ of the former (Mamalakis et al., 2021). Regarding the functional form of $F$, Mamalakis et al. (2021) noted that the function $F$ can be of any arbitrary choice (depending on what type of network the analyst wants to benchmark, e.g., a fully connected, a CNN, etc.), as long as it has such a form so that the attribution of any output to the corresponding input is objectively derivable.

We herein consider an idealized classification task that is specifically designed for CNN applications and is inspired by remote sensing tasks in geosciences where spatial patterns (such as cloud objects, weather fronts, etc.) need to be tracked and extracted (see e.g., Hilburn et al., 2021). We generate a series of inputs that consist of 2D (single channel) images where circular and square frames are present, and we task the CNN to classify each image depending on which class of frames covers more area (i.e., has more pixels). More specifically, each input image consists of 65×65 pixels (i.e., a total of $d = 4225$ pixels), with the input features being binary variables, $\mathbf{X} \in \{0,1\}^d$. If a pixel $i$ for a sample $n$ belongs to a circular or a square frame then $x_{i,n} = 1$ (i.e., see for example the dark red pixels of the square frame in the top left plot of Figure 1), while $x_{i,n} = 0$ otherwise; see Figure 1 for some examples of the synthetic input images. The number of frames per class, the size of the frames, and the positioning of the frames is random in each image, and no frame overlap is allowed to occur. In terms of the output of the dataset, two separate classes that all input images are classified into are defined: Class 1: the circular frames in the image cover more area than the square frames. Class 2: the square frames in the image cover more area than the circular frames. Thus, the synthetic output of the dataset is a series of logical values indicating which class each input image corresponds to, and the network is trained to classify the images between the two classes.

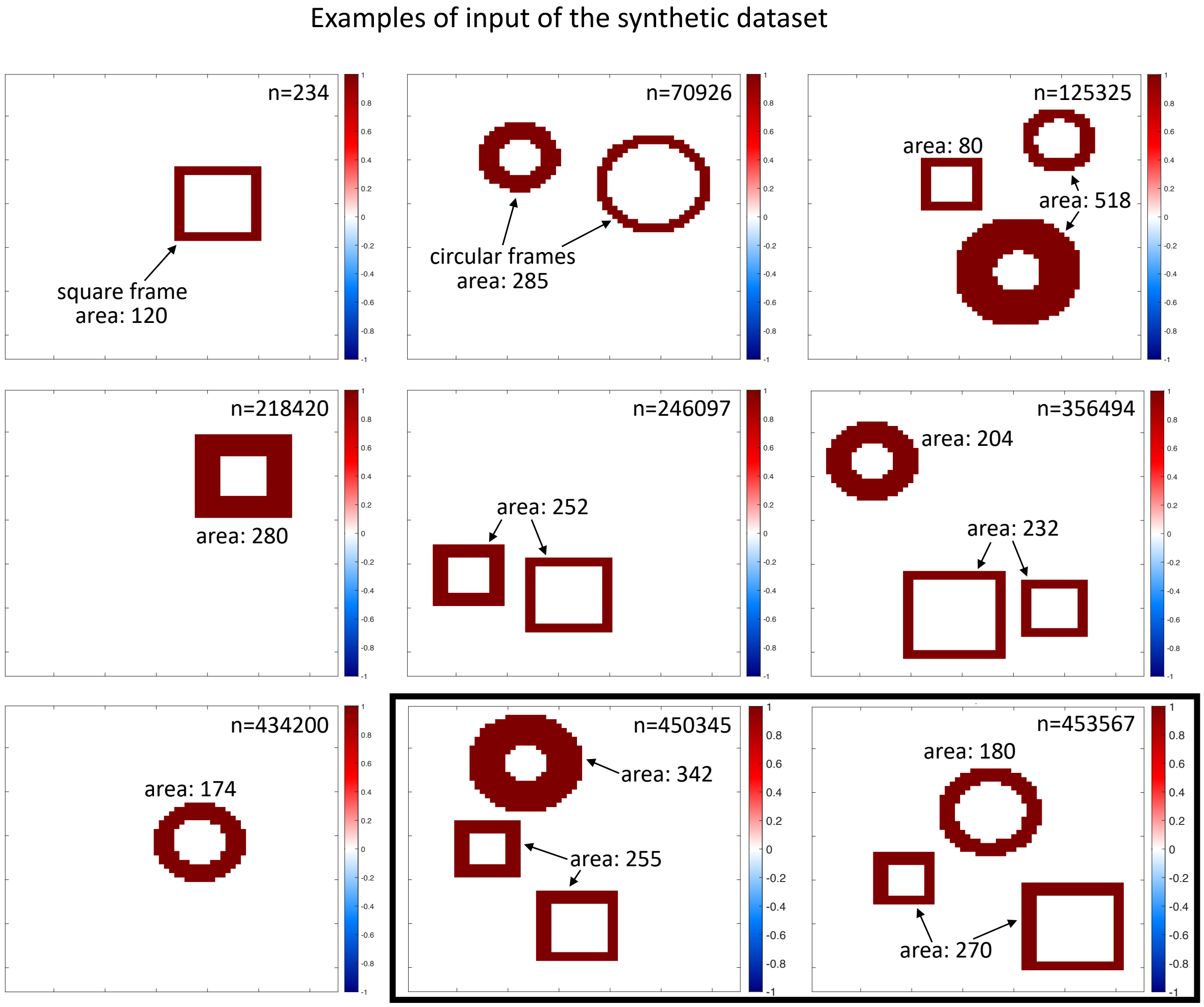

**Figure 1.** Examples of input of the synthetic attribution benchmark dataset. Details about this synthetic dataset are provided in section 2.1. In many cases, the answer as to which class of frames covers more area is easy to get simply with visual inspection. However, there are also cases where the answer is more difficult to disentangle (such as sample #356494 or #450345). The testing performance of the trained CNN (see architecture in Figure 2a) was slightly above 99% accuracy; higher performance than what a human eye would do. The examples highlighted in the black box are from the testing dataset and are analyzed further in section 3 (see Figures 3-4).

By choosing this simple, idealized classification task we achieve three things. First, for any model to be able to correctly classify these synthetic images, it needs to be able to extract spatial patterns of different shapes. This makes a CNN (our study's focus) be the most suitable type of network to address this classification task (LeCun et al., 2015). Second, the simplicity of the current task makes it possible for us to objectively derive the ground truth of the attribution: pixels in an image that belong to any circular (square) frame contribute positively (negatively) to the probability that class 1 is true and negatively (positively) to the likelihood of class 2; note that this is valid when considering a blank image as the baseline. Third, an immediate consequence of the latter rule of attribution is that, as we will see in Figures 3-6, there are many cases where both positive and negative contributions appear in the same explanation. This means that with this benchmark, we can assess which XAI methods can disentangle the sign of the contribution of specific input patterns to the output, an aspect that is often overlooked (Kohlbrenner et al., 2020). In summary, this synthetic dataset fits our current scope to objectively assess XAI for CNN applications; for the connection of this dataset with the mathematical framework of *additively separable functions* introduced by Mamalakis et al. (2021) see Appendix A.

The CNN that we use for this classification task consists of three pairs of convolutional and max pooling layers followed by three fully connected layers (see panel (a) in Figure 2). The output layer consists of two neurons, with the first (the second) calculating the likelihood that the circular (square) frames cover more area. We use ReLU activations in all layers apart from the output layer, where we use the softmax function. The CNN is trained using 450,000 samples, while 50,000 samples are used for testing. The testing classification accuracy is slightly above 99%, i.e., less than 1% of the testing images are misclassified by the network. The reason that we chose to generate such an unrealistically high sample size is so that the CNN can learn almost perfectly the underlying function $F$. Only under this condition is it fair to use the ground truth of attribution as a benchmark for the XAI methods, since any deviation between the two should mostly arise from XAI limitations and to a lesser degree from poor training of the network. However, we note that discrepancies between XAI output and the ground truth shall always exist due to the fact that the CNN is a close approximation (not identical) to the function $F$.

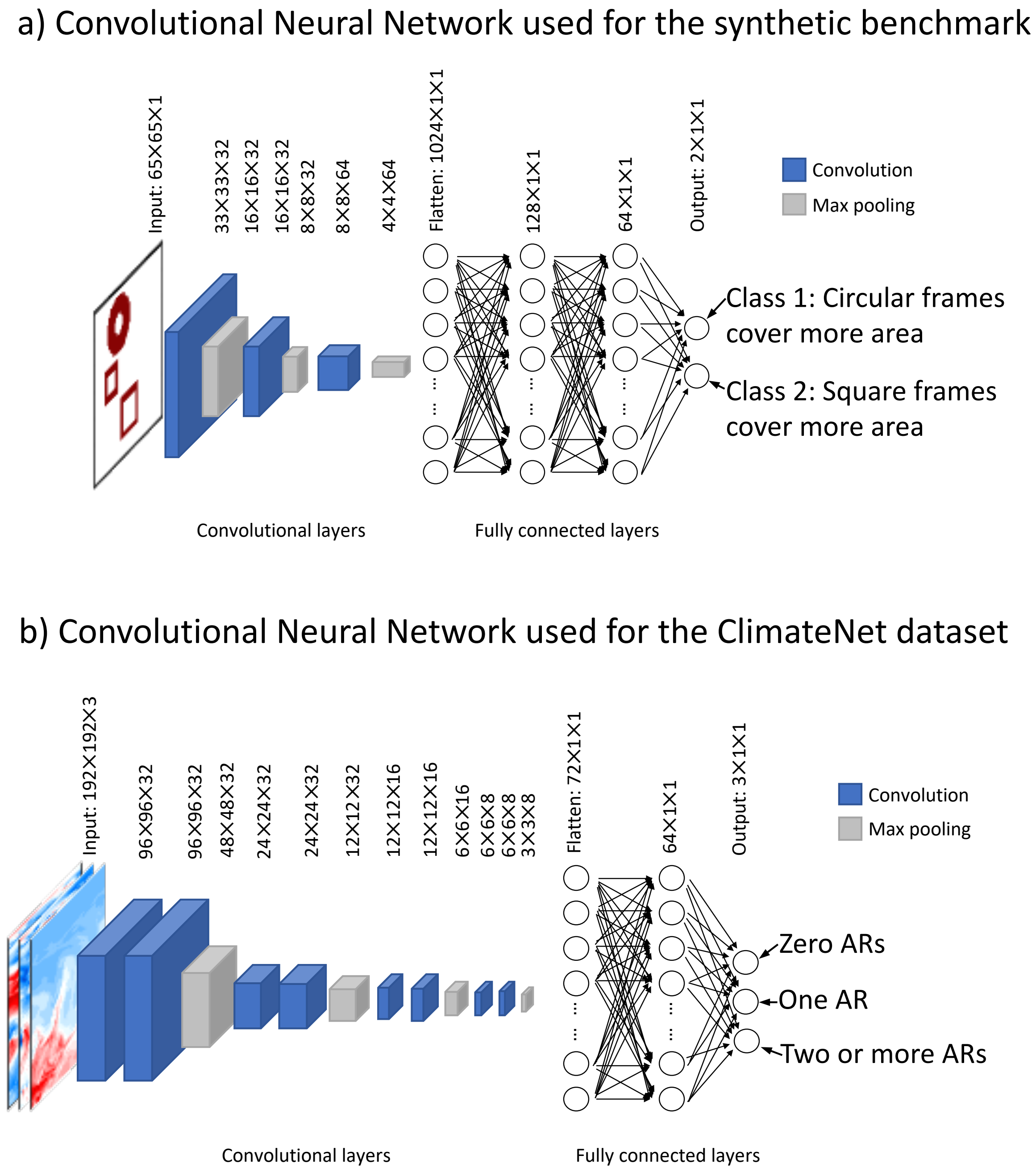

**Figure 2.** Specific architectures of the Convolutional Neural Networks that were used in the two classification problems of our study.

### 2.2. ClimateNet dataset

As a second application, we employ a more climate-related task where there is no ground truth of the explanation available (as is the case in most geophysical studies). This second task aims to validate the insights about XAI gained from the first task in a more climate-related setting. For our second classification problem we use the ClimateNet dataset (Prabhat et al., 2021). The ClimateNet dataset is a publicly available dataset (https://portal.nersc.gov/project/ClimateNet/) that consists of daily outputs of climate simulations from the Community Atmospheric Model (CAM5.1). Each daily output includes snapshots of many different variables like precipitation, vertically integrated precipitable water and temperature and wind velocities at different pressure levels. Also, for each simulated day in the ClimateNet dataset a labeled world map is available, where expert meteorologists and scientists have detected the locations over which atmospheric rivers (narrow elongated bands of enhanced water vapor in the atmosphere; ARs) and tropical cyclones occur around the world on that specific day. This labelling has currently been done for 456 days of the simulated historical years 1996-2013, which is the total sample size of the dataset (see an example of a simulated day in Figure S1).

We build a CNN to classify these daily snapshots from the ClimateNet dataset in terms of how many ARs occur on the corresponding day. More specifically, we use a 3-channel image as our input with zonal and meridional wind velocities at 850mb pressure level and vertically integrated precipitable water constituting the three channels. Based on the expert labeling that is available in the dataset, the CNN is then trained to classify the input into three different classes: zero, one, and two or more ARs occurring on the corresponding simulated day. The architecture of the CNN consists of four sets of two convolutional layers and one max pooling layer followed by two fully connected layers (see panel (b) in Figure 2). We use ReLU activations in all layers of the network apart from the output layer, where we use the softmax function. The output layer consists of three neurons, with the first, second and third neuron computing the likelihood that zero, one, and two or more ARs occur on the simulated day, respectively. Because a size of 456 samples is small to train and test a deep CNN, we cut each snapshot in the dataset into six equally-sized segments (three segments in each hemisphere; see an example in Figure S1). We use simulations in years 1996-2010 for training and in years 2011-2013 for testing, where each input channel is standardized by using the all-sample and all-pixel mean and standard deviation of the corresponding variable. The end result of this preprocessing is that 2,370 samples were used for training, 366 samples were used for testing, and the input image consists of 192×192×3 pixels. The trained CNN exhibits a classification accuracy of 62% for the testing data.

This climate-related classification task is fairly similar to the idealized classification task of the previous section in that both tasks require the adopted model to learn to extract (and compare or neglect) specific spatial patterns. In the idealized dataset, the model is required to learn to compare the area of two different classes of spatial patterns (square and circular frames), while in the ClimateNet dataset, the model is required to learn to extract spatial patterns that resemble ARs but to neglect all other spatial patterns that might be present (e.g., tropical cyclones). This similarity between the two problems allows us to validate the XAI insights that are gained from the idealized task in a similar but more climate-related second task.

### 2.3. XAI methods

For our assessment, we consider some of the most popular XAI methods for CNNs that have been proposed in the computer science literature. To keep this section as concise as possible, we only briefly describe how each method explains the network in the following list (the category that each method belongs to is provided in parenthesis; see also Table 1). For more details on the methods' analytical formulas, the reader is referred to Appendix B and the corresponding studies cited below.

**Gradient (sensitivity):** This method (Simonyan et al., 2014) assesses the importance of the input features based on the *sensitivity*. Sensitivity refers to how much the value of the output will change for a unit change in a specific feature and is estimated here by the first partial derivative of the network's output with respect to the feature.

**Smooth Gradient (sensitivity):** This method (Smilkov et al., 2017) also computes the gradient, but it does so by averaging the gradients over a perturbed number of inputs with added noise. This aims to increase the robustness of the results (i.e., reduce the noise).

**Input*Gradient (attribution):** This method (Shrikumar et al., 2016; 2017) assesses the *attribution* of the output to the input (see detailed differences between *sensitivity* and *attribution* in Appendix C). Attribution refers to the marginal contribution of an input feature to the output and is estimated here by multiplying (pixel-wise) the input with the gradient.
**Integrated Gradients (attribution):** This method (Sundararajan et al., 2017) uses a reference vector (e.g., for which the network's output is zero). It then estimates the contribution of each feature as the product of the average of the gradients at points along the straightline path from the reference point to the input with the distance of that path. Integrated Gradients is similar to Input*Gradient but is designed to account for nonlinearities in the model that is being explained.
**Layer-wise Relevance Propagation (LRP; attribution):** This method (Bach et al., 2015) propagates the network's output back to neurons of lower layers, until the input layer is reached. In the back propagation phase the relevance/importance of each neuron to the output is estimated, based on different propagation rules. In this study, we consider the most popular LRP rules: i) The $\mathbf{LRP_z}$ rule (Bach et al., 2015), which distributes the relevance of each neuron based on the values of the localized preactivations that are directed to it. ii) The $\mathbf{LRP_{\alpha1\beta0}}$ (Bach et al., 2015), which is similar to $LRP_z$ but considers only positive preactivations. iii) The $\mathbf{LRP_{comp}}$ (Kohlbrenner et al., 2020), which combines the two previous rules; it applies the $LRP_z$ rule to distribute the relevance in the fully-connected layers of the CNN and the $LRP_{\alpha1\beta0}$ rule in the convolutional layers. iv) The $\mathbf{LRP_{comp/flat}}$ (Bach et al., 2016; Kohlbrenner et al., 2020), which is similar to $LRP_{comp}$ but additionally applies a flat rule in the very lowest layer(s). The flat rule distributes relevance uniformly to all connected neurons, without considering the preactivations values.
**Deep Taylor Decomposition (attribution):** This method (Montavon et al., 2017) applies a local Taylor decomposition, to decompose each neuron's relevance to the neurons of the lower layer. It is applied recursively until the importance of the input features is obtained. Deep Taylor is equivalent to $LRP_{\alpha1\beta0}$ for networks that use ReLU activations.
**PatternNet (signal)** and **PatternAttribution (attribution):** These methods (Kindermans et al., 2017a) are based on the idea that every input image consists of a signal component (all of the information in the input that is relevant to the prediction task) and a distractor (all of the distracting information that is irrelevant to the prediction task). The method PatternNet performs a layer-wise back-projection of the signal to the input space. In each layer, the signal is approximated as a superposition of neuron-wise, local signal estimators. This is done recursively, until the signal of the network's output in the input image is estimated. PatternAttribution aims to estimate the attribution of the network's output to the input (i.e., not simply the signal), by applying the same layer-wise back-projection approach, but also considering the weight vector that connects subsequent layers.
**Deep SHAP (attribution):** This method (Lundberg and Lee, 2017) approximates the Shapley values (originally discovered in the field of the cooperative game theory; Shapley, 1953) for the entire network by computing the Shapley values for smaller components of the network and propagating them backwards until the input layer is reached (similar in philosophy to LRP, PatternNet and PatternAttribution). Shapley values have been shown to satisfy desired properties regarding the explanation (e.g., local accuracy, missingness and consistency; Lundberg and Lee, 2017), which is not necessarily the case with other XAI methods (e.g. LRP, Input*Gradient, etc.).

## 3. Results

In this section, we present the results of applying the XAI methods first to the synthetic dataset and then to ClimateNet. We highlight that for the synthetic dataset, methods Gradient, Smooth Gradient and PatternNet are not directly comparable to the derived ground truth of attribution, since they estimate the *sensitivity* or the *signal* of the output to the input; rather than the *attribution* of the output to the input (see Appendix C for the difference between *sensitivity* and *attribution*). However, they are included in the intercomparison due to their popularity and for the sake of completeness.

**Table 1.** Summary of XAI methods considered in this study. Practical strengths (✓) and weaknesses (×) of each method are also reported.

| XAI method | | Brief summary of the method | Desired property for CNN applications as explored in this study | | | Extra comments/insights |
|---|---|---|---|---|---|---|
| | | | disentangles the sign of relevance | insensitive to gradient shattering | not ignorant to zero input | |
| **Gradient** (Simonyan et al., 2014) | | Calculates the first partial derivative of the model output with respect to the input. (sensitivity) | ✓ | × | ✓ | Estimates the sensitivity of the output to the input, which is not the same as the attribution; see Appendix C |
| **Smooth Gradient** (Smilkov et al., 2017) | | Calculates the average gradient across many perturbed inputs. (sensitivity) | ✓ | × | ✓ | |
| **Input*Gradient** (Shrikumar et al., 2017) | | Multiplies the input with the gradient. (attribution) | ✓ | × | × | |
| **Integrated Gradients** (Sundararajan et al., 2017) | | Multiplies the average gradient along the straight line between the input point and a reference point with the corresponding distance between the two points. (attribution) | ✓ | × | ✓ | |
| **LRP** | **α1β0** (Bach et al., 2015) | Layer-wise back propagation of each neuron's relevance based on the α1β0-rule. (attribution) | × | ✓ | × | Considers only positive preactivations |
| **LRP** | **z** (Bach et al., 2015) | Layer-wise back propagation of each neuron's relevance based on the z-rule (attribution) | ✓ | × | × | Equivalent to Input*Gradient for networks using ReLU activations |
| **LRP** | **comp** (Kohlbrenner et al., 2020) | Layer-wise back propagation of each neuron's relevance by combining the α1β0-rule and the z-rule. (attribution) | ✓ | ✓ | × | Combines the strengths of LRPz and $LRP_{\alpha1\beta0}$ |
| **LRP** | **comp/flat** (Kohlbrenner et al., 2020) | Layer-wise back propagation of each neuron's relevance by combining the α1β0-rule, the z-rule and the flat rule. (attribution) | ✓ | ✓ | ✓ | Provides a coarser picture of attribution; not suitable if local accuracy necessary |
| **Deep Taylor** (Montavon et al., 2017) | | Applies Taylor decomposition of the relevance function for each neuron recursively. (attribution) | × | ✓ | × | Equivalent to $LRP_{\alpha1\beta0}$ for networks using ReLU activations; not defined for negative predictions |
| **PatternNet** (Kindermans et al., 2017a) | | Calculates the signal in the input for each neuron recursively. (signal) | × | ✓ | ✓ | Estimates the signal (not the same as the attribution) |
| **PatternAttribution** (Kindermans et al., 2017a) | | Calculates the attribution in the direction of the signal for each neuron recursively. (attribution) | × | ✓ | ✓ | |
| **Deep SHAP** (Lundberg and Lee, 2017) | | Approximates Shapley values for each neuron recursively (attribution) | ✓ | × | ✓ | Based on well-founded theory; computationally expensive |

### 3.1. Synthetic dataset

In Figures 3-4, we explore the CNN strategy for two samples from the synthetic dataset. In both samples, two square frames and one circular frame are present. In Figure 3, the square frames cover more area (specifically 270 pixels versus 180), and in Figure 4, the circular frame covers more area (specifically 342 pixels versus 255), while in both cases, the CNN has correctly classified the input images. In terms of the ground truth of attribution in Figure 3, we expect that the pixels of the square frames increased the certainty of the network (i.e., increased the likelihood of class 2), while the pixels of the circular frame decreased it (this is valid when considering a blank image as our baseline). That is, if it was not for the circular frame, the certainty of the model would have been higher. The opposite is true in Figure 4.

As is evident in both figures, despite all methods being applied to explain the same exact prediction, different XAI methods lead to different explanations. Specifically, despite most methods identifying the frames as important features, some methods exhibit relative noisier results, and there is no consensus regarding the sign of the attribution. If this was a classification problem we knew nothing about (as could be the case for a typical geoscience setting), it would be difficult to reach certain conclusions about the decision strategy of the network. However, by knowing the ground truth of attribution in these examples, we can assess the fidelity of each of the methods and also understand the lack of consensus in the results.

First, Gradient is shown to produce somewhat noisy patterns. For shallow networks, some studies suggest that the gradient resembles a Brownian motion and exhibits spatial coherence, while for deeper networks the gradient converges to white noise and the spatial autocorrelation vanishes (Balduzzi et al., 2017). This phenomenon is known in the computer science literature as “gradient shattering” (Balduzzi et al., 2017). Although our network is not very deep (less than 10 layers), the noise in the results of the gradient can be partially attributed to gradient shattering. Despite this, one can see that the square (circular) frames are highlighted with mostly positive (negative) values in Figure 3, while opposite results are shown in Figure 4, which is consistent with what we expect in both cases. Moreover, in both figures, the gradient vanishes away from the frames. This means that the CNN has correctly learned that if one were to increase the value of any pixel away from the frames this would not affect the chances of either class, because isolated pixels constitute neither a circular nor a square frame. Smooth Gradient produces quite different results, namely mostly negative gradients in Figure 3 and mostly positive gradients in Figure 4.

Results from the Input*Gradient and Integrated Gradients methods are very similar and close to the ground truth of attribution (pattern correlation with the ground truth is on the order of 0.5-0.6 in all examples). In Figure 3, the square (circular) frames are highlighted with mostly positive (negative) attributions, while in Figure 4, we obtain the opposite results. Pixels outside of the frames receive zero attribution. However, both methods may suffer from the effects of gradient shattering in the same way as Gradient, since they are directly connected to the latter (see Eq. B.3-B-4 in Appendix B). Indeed, as we can see in Figures 3-4, attributions exhibit some level of noise, which for a deeper network might be so high that it can severely limit comprehensibility (e.g., see Figure 7).

The PatternNet method correctly highlights all three frames (as well as some pixels away from the frames) as important in containing information (a signal) for the decision of the CNN. PatternAttribution correctly highlights the two square frames in Figure 3 and the circular frame in Figure 4 as contributing positively to the CNN’s decision. However, PatternAttribution does not very effectively distinguish between positive and negative contributions in either example, because in both cases, it assigns positive attribution to the frames that are actually contributing negatively to the CNN’s decision. Its pattern correlation with the ground truth is on the order of 0.4 in both examples.

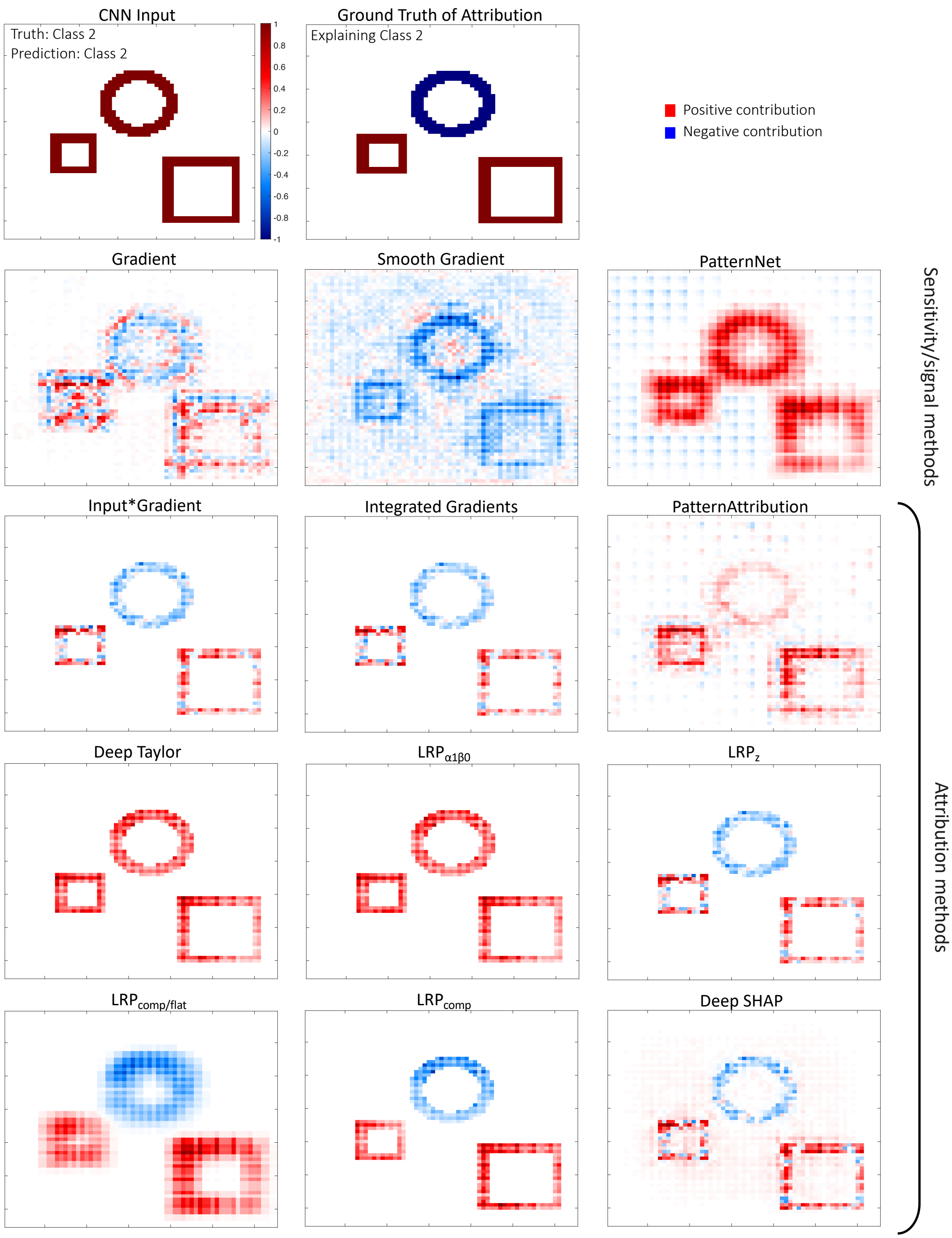

**Figure 3.** Explanations from different XAI methods of the strategy of the CNN for the synthetic dataset and sample #453567. The CNN has successfully classified this image to class 2, i.e., the square frames cover more area (area: 270) than the circular frames (area: 180). XAI methods are applied to explain the successful prediction. For each heatmap, we divided all values by the maximum (in absolute terms) value. The ground truth of attribution is derived using a blank image (image with zeros) as a baseline.

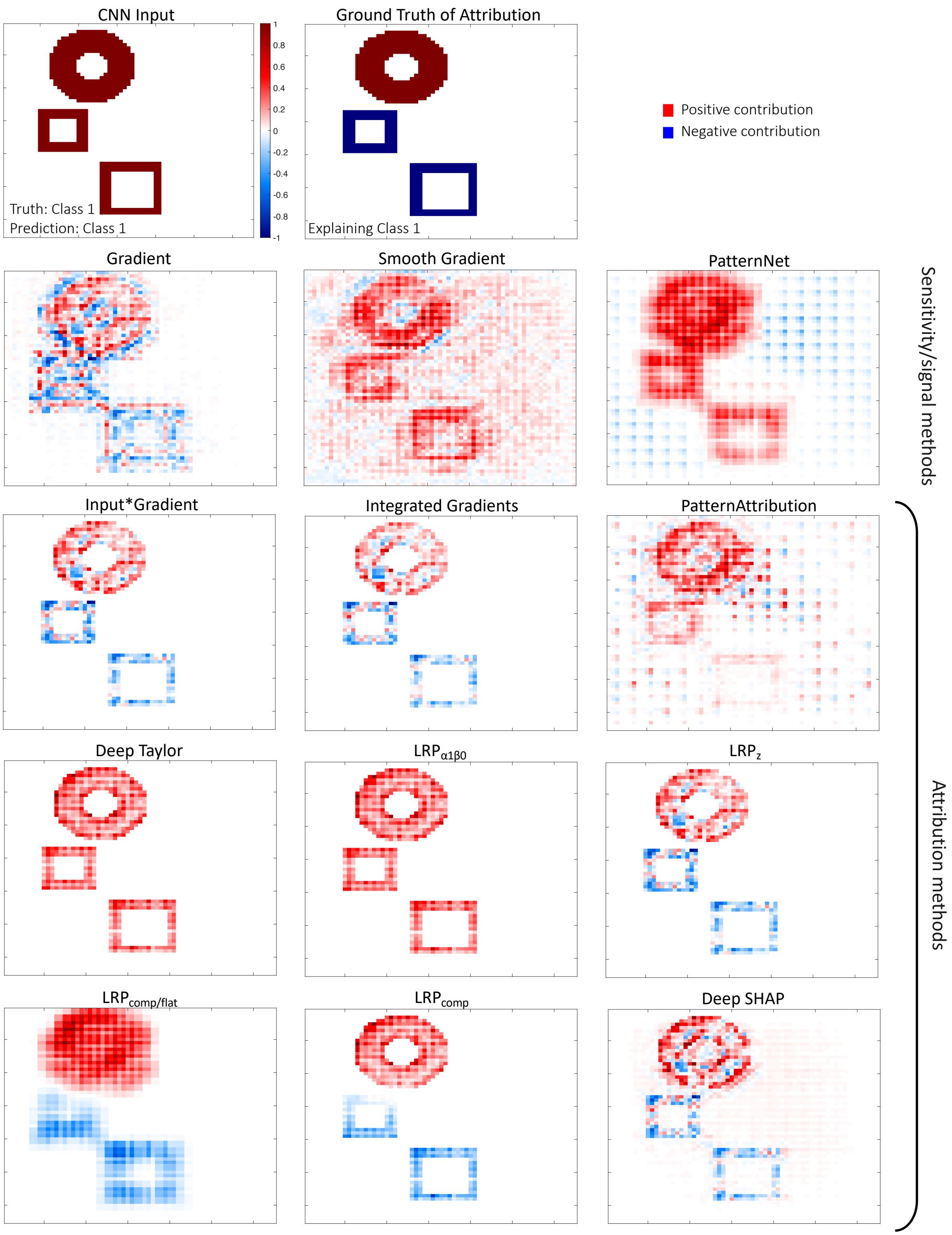

**Figure 4.** Same as in Figure 3, but for the sample #450345. The CNN has successfully classified this image to class 1, i.e., the circular frames cover more area (area: 342) than the square frames (area: 255). XAI methods are applied to explain the successful prediction.

The results of Deep Taylor and the $LRP_{\alpha 1 \beta 0}$ rule are identical, since these two methods are equivalent for networks with ReLU activations (Samek et al., 2016; Montavon et al., 2017). Both methods pick up the corresponding three frames, and pixels outside of the frames receive zero attribution. However, all frames receive positive attributions in both figures, which is not consistent with the ground truth of attribution (the results of these two methods exhibit a correlation with the ground truth of only about 0.1-0.2). It has recently been noted that $LRP_{\alpha 1 \beta 0}$ propagates the sign of the before-softmax value back to the input (e.g., Kohlbrenner et al., 2020), and thus, it is not able to distinguish between positive and negative contributions of different features[1]. Due to this property, $LRP_{\alpha 1 \beta 0}$ is known to provide smoother (not very noisy) results compared to other LRP rules, however with limited local accuracy, since the negative preactivations are not being considered in this rule (see Eq. B.6 in Appendix B). Results from $LRP_z$ are the same as those from the Input*Gradient since the two methods are equivalent when explaining networks that use ReLU activations (Ancona et al., 2018; 2019).

The results from $LRP_{comp}$ seem to be the most consistent and very similar to the ground truth of attribution (this method exhibits the highest correlation with the ground truth on the order of 0.8-0.9). As mentioned in the previous section, this method combines $LRP_{\alpha 1 \beta 0}$ and $LRP_z$ in an attempt to get the best from both rules: as shown in Figures 3-4, it is able to maintain local accuracy, and thus, distinguish between positive and negative contributions (owing to the use of $LRP_z$), while at the same time returning smooth results, thus, eliminating the effect of gradient shattering (owing to the use of $LRP_{\alpha 1 \beta 0}$). The rule $LRP_{comp/flat}$ is shown to provide a coarser but similar picture of attribution to the $LRP_{comp}$ (correlation with the ground truth on the order of 0.7). This verifies arguments in previous studies (Bach et al., 2016) that if the analyst/scientist is not interested in local accuracy, but they only need to obtain a coarse picture of the attribution, this is a suitable rule to use. Lastly, the method Deep SHAP is shown to provide attributions that are close to the ground truth, but relatively noisier (correlation with the ground truth on the order of 0.5-0.6). Results are similar to the results of Input*Gradient, Integrated Gradients and $LRP_z$.

In Figure 5, we repeat the results of Figure 4, but now, we aim to detect which features in the input made the CNN assign a very small probability to class 2. We note that in geoscientific applications, it is always good practice to use XAI to explain not only the predicted class, but also the rejected class(es), since this may provide further insight. The ground truth of attribution in Figure 5 shows the opposite of what shown in Figure 4: the pixels of the square frames increase the likelihood of class 2, while the pixels of the circular frame decrease it. The XAI results verify most of the arguments made in the discussion of the previous figures. First, methods like Gradient, Input*Gradient, Integrated Gradients, $LRP_z$, Deep SHAP etc. are able to disentangle the sign of the attribution but might be partially affected by gradient shattering. The $LRP_{\alpha 1 \beta 0}$ rule provides smooth results but cannot disentangle the sign of the attribution. Specifically, it assigns negative attributions to all frames (similarly to PatternNet and PatternAttribution), because the before-softmax value that corresponds to class 2 is a negative number in this example. Deep Taylor does not return any results, since this method is only defined for positive network outputs (Montavon et al., 2017). Lastly, the method $LRP_{comp}$ is again shown here to provide the most consistent attribution compared to the ground truth, as it is able to provide smooth results and also disentangle the sign of the attribution.

To explore the above insights more quantitively and across many samples we have calculated the distribution of the correlation with the ground truth for each of the XAI attribution methods (not shown). This analysis showed that $LRP_{comp}$ exhibits systematically the strongest correlation with the ground truth, $LRP_{\alpha1\beta0}$ exhibits the weakest correlation, while the rest of the methods fall in between, similar to what Figures 3-5 suggest.

---

[1] This can be explained easily by looking at the formula in Eq. (B.6) and setting $\alpha = 1$ and $\beta = 0$ to obtain the $LRP_{\alpha 1 \beta 0}$ rule: Because the ratio $\frac{{z_{ij}}^+}{{z_j}^+}$ is by definition a positive number, then the relevance of any neuron in the lower layer $R_i^{(l)}$ has the same sign as the relevance of the neuron in the upper layer $R_j^{(l+1)}$, and this sign is maintained and recursively propagated back to the input layer. Thus, when the before-softmax value of the class that is being explained is a positive (negative) number then the corresponding heatmap will show only nonnegative (nonpositive) values.

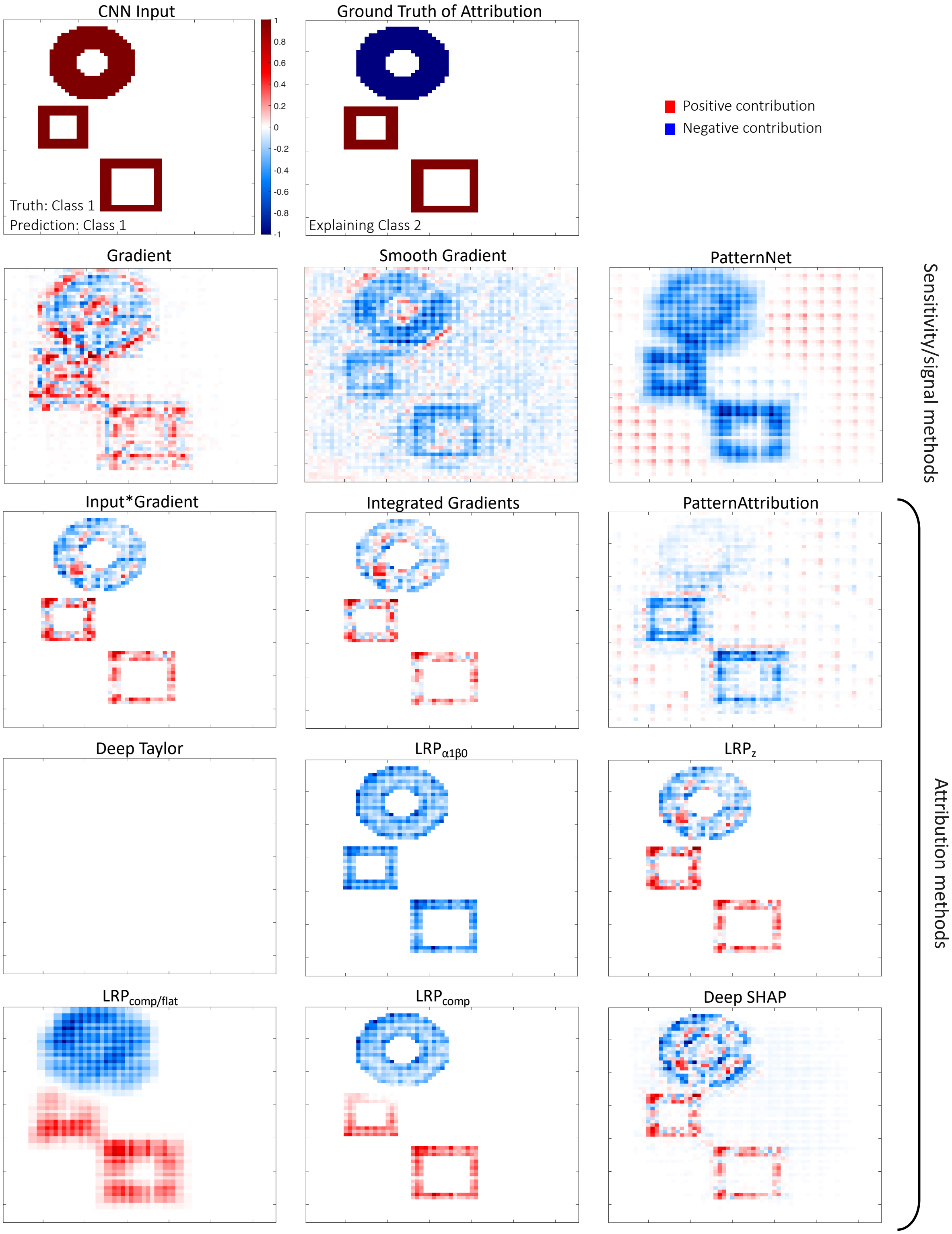

**Figure 5.** Same as in Figure 4, but here XAI methods are applied to explain why the CNN correctly predicted that the class 2 is not true (i.e., explaining the low probability that the CNN assigned to class 2).

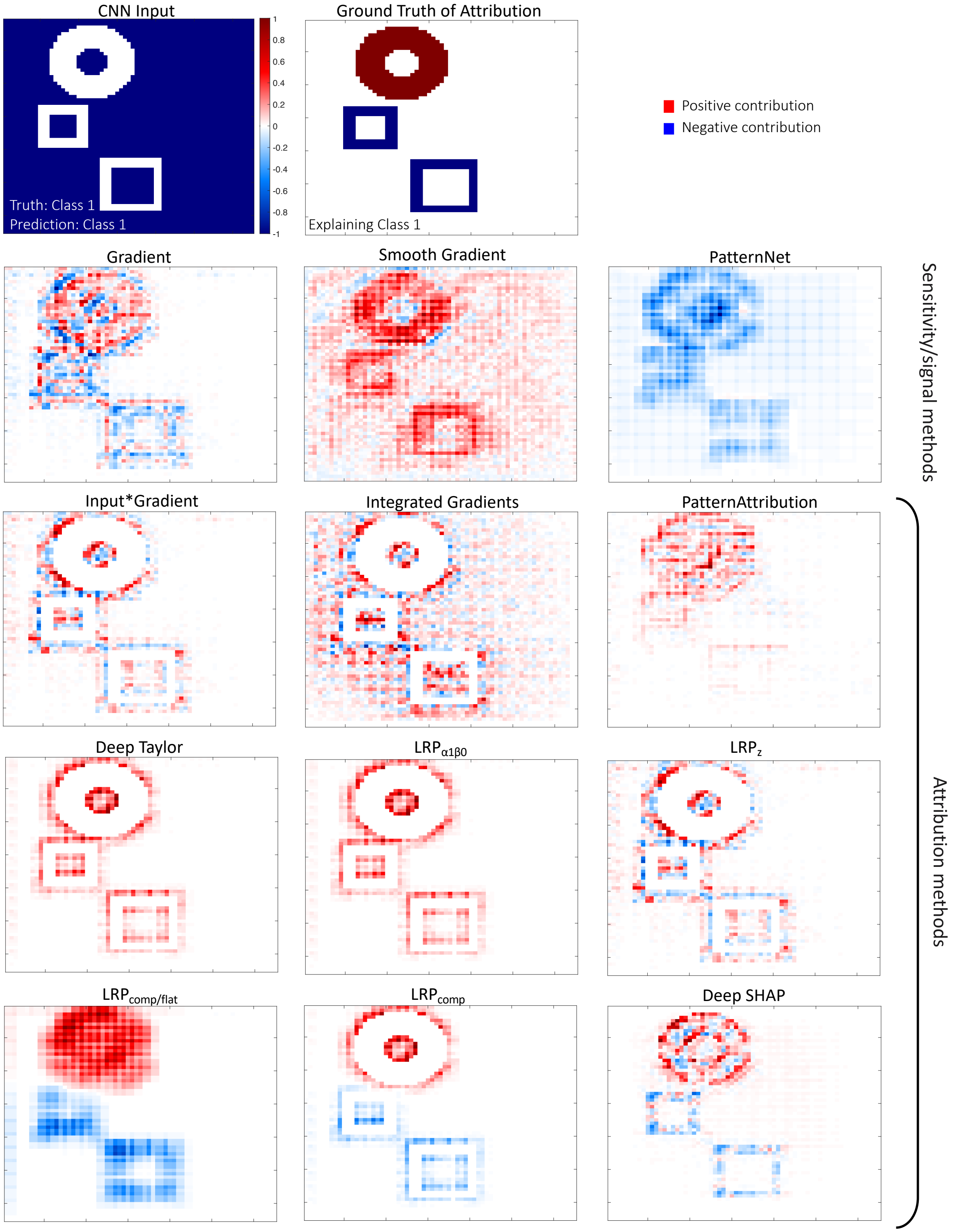

**Figure 6.** Same as in Figure 4, but after a shift of -1 has been applied to the input. Also note that the ground truth of attribution is derived using a baseline image with all feature values equal to -1.

Next, we explore the sensitivity of the XAI results to input transformations. In geoscientific applications, input transformations may represent modifications of the units of an input variable (e.g., from degrees Kelvin to Celsius) or the scaling (anomalies about zero vs raw measurements), thus, it is of high importance to investigate their effect. To do so, we perform the following experiment that is inspired by Kindermans et al. (2017b): we apply a uniform shift of $s = -1$ to all pixels in all input images of the synthetic dataset. The features of the shifted input are binary variables with $\mathbf{X}^* \in \{-1,0\}^d$: the pixels of the frames are equal to 0 and the non-frame pixels are equal to -1. We then consider the already trained CNN and simply change the biases of its first layer to account for the shift in the input: for any $j$ neuron in the first hidden layer the new bias term is modified as $b_j^* = b_j - \sum_i w_{ij} s = b_j + \sum_i w_{ij}$. With this modification, the predictions of the modified CNN (denoted CNN*) when using the shifted input (denoted $\mathbf{X}^*$) are the same as those of the CNN in the original setting[2].

In Figure 6, we apply XAI methods to explain the decision strategy of this modified CNN* for the same prediction as in Figure 4. The methods Gradient, Smooth Gradient, PatternAttribution, $\text{LRP}_{\text{comp/flat}}$ and Deep SHAP provide similar results, which makes them "input shift invariant" (Kindermans et al., 2017b). The reason for the invariance in their results is: i) The gradient of a constant is zero, so methods Gradient and Smooth Gradient are expected to be "input shift invariant". ii) The method $\text{LRP}_{\text{comp/flat}}$ applies a flat rule in the lowest layers, which distributes relevance uniformly to any input feature that is connected to a neuron in the upper layer, without considering the value of preactivations (see Eq. B.7). Thus, since the architecture of the modified CNN* and all preactivations in all layers except the lowest one are the same as in the original setting, the feature attributions are the same. iii) Both PatternAttribution and Deep SHAP use the range of variability of the input features in the training dataset in order to assess feature importance, thus, the input shift is taken into account.

In contrast, Input*Gradient, Integrated Gradients, and all the rest of LRP rules show very different results with the shifted input compared to Figure 4, with most of these methods highlighting the perimeter of the three frames while the body of the frames receives zero attribution. This indicates a sensitivity of these methods to input transformations. In our example, this sensitivity originates from the fact that these methods are theoretically unable to assign attribution to a zero value in the input (i.e., the body of the frames in Figure 6). Indeed, the formulas in Eq. (B.3)-(B.4) in Appendix B show that Input*Gradient and Integrated Gradients (when using a blank image as reference) always assign a zero attribution to a zero input by construction. Similarly, all LRP rules except $\text{LRP}_{\text{comp/flat}}$ perform the relevance re-distribution based on the preactivation value $w_{ij}x_i$, thus zero inputs automatically receive a zero attribution. For the rest of the manuscript, we will refer to this systematic behavior of assigning zero attribution to a zero input and ignoring the impact it could have to the network's output as the "ignorant to zero input" issue. Input*Gradient, and all LRP rules except $\text{LRP}_{\text{comp/flat}}$ are "ignorant to zero input". The "ignorant to zero input" issue did not show up in Figures 3-4 (if anything, it worked to the advantage of these methods), since pixels with a zero value were expected to receive zero attribution in those examples. In general however, this issue can provide a distorted picture of the decision strategy of the CNN. A clear example is Figure 6, where according to Input*Gradient, and most LRP rules, the frames are not important to the CNN decision[3]. We also note that if we train a completely new CNN to classify the shifted images and use XAI to explain its predictions (see Figure S2), we observe very similar results with Figure 6, which further verifies the validity of the above remarks.

---

[2] Any activation value of the neurons in the first hidden layer $x_j^*$ is equal to the corresponding activation in the original setting: $x_j^* = \text{ReLU}\left(\sum_i w_{ij} x_i^* + b_j^*\right) = \text{ReLU}\left(\sum_i w_{ij}(x_i - 1) + b_j + \sum_i w_{ij}\right) = \text{ReLU}\left(\sum_i (w_{ij} x_i - w_{ij}) + b_j + \sum_i w_{ij}\right) = \text{ReLU}\left(\sum_i w_{ij} x_i + b_j\right) = x_j$; see also Kindermans et al. (2017b).

[3] As a second example, let us consider we wanted to explain the prediction of a (supposedly perfectly trained) network that simulates the function $F(\mathbf{X}) = \sum_{i=1}^{d} \cos(\mathrm{X}_i)$ at the point $\mathbf{x} = \mathbf{0}$. An "ignorant to zero input" method

The results of this section highlight three important issues of XAI methods, namely, the effect of gradient shattering, the issue of disentangling the sign of the attribution, and the "ignorant to zero input" issue (see Table 1). All of these issues may limit the user's understanding of the decision-making strategy of a CNN and no method was shown to be optimal.

### 3.2. ClimateNet

In Figures 7-8, we apply the same XAI methods to explain CNN predictions for the ClimateNet dataset. For this dataset, there is no clearly defined ground truth for the attribution of the output to the input. Even though the dataset contains labeled maps by experts (i.e., Figure S1d), these cannot act as a ground truth for the attribution, as the CNN may employ patterns or climate information outside of the regions of the ARs for making its predictions. Thus, we cannot assess the XAI fidelity for this application as we did for the synthetic dataset. Instead we use this dataset to examine whether, and how, the properties and artifacts of different XAI methods that were identified in the previous section manifest in a more climate-related prediction setting. Furthermore, by providing this example we seek to illustrate how the knowledge of relative strengths and weaknesses of each XAI method affects our interpretation of the corresponding XAI results.

In the specific sample that we consider, two ARs have been detected by the expert scientists, and the CNN correctly assigned this input to the class of two or more ARs. In Figure 7, we present the XAI results that explain which features in the first channel of the input image (the zonal wind at 850mb pressure level; U850) the CNN used to make this prediction. Similar to the previous dataset, the obtained results are very different when using different XAI methods, which makes the interpretation of the decision-making strategy of the CNN challenging. First, in accordance to the remarks of the previous section, one can see that the results of the methods Gradient, Smooth Gradient, Input*Gradient, Integrated Gradient and $LRP_z$ are very noisy (Figure 7), and based on these methods one cannot make any robust inferences about the CNN's strategy. For ClimateNet, the CNN that we use is almost twice as deep as in the previous dataset (see Figure 2), and thus, the gradient shattering has a detrimental effect on the explanations.

Focusing on the rest of the methods, PatternNet highlights all features in the input where zonal wind is positive, indicating that these features contain important information for the network. PatternAttribution seems to primarily highlight one of the two wind patterns that are associated with the two ARs. The methods Deep Taylor and $LRP_{\alpha 1 \beta 0}$ provide only positive attributions to all highlighted features (recall here from the previous section that these methods do not disentangle the sign of the attribution), and they assign the highest attribution to the two positive wind patterns that are associated with the ARs. The same features are highlighted more clearly when using the methods $LRP_{comp}$ and Deep SHAP. These two methods are relatively more insightful, since: i) $LRP_{comp}$ is a "best practice" implementation of LRP (Kohlbrenner et al., 2020) and it combines the strengths of the rules $LRP_{\alpha 1 \beta 0}$ and $LRP_z$, and ii) Deep SHAP has been proven to satisfy desirable properties of consistency, local accuracy and missingness (Lundberg and Lee, 2017), it successfully disentangles the sign of the attribution and does not exhibit the "ignorant to zero input" issue. Last, the results from $LRP_{comp/flat}$ show how the attribution is distributed when one

---

would assign a zero attribution to all input features, just because $x_i = 0, \forall i$. This ignores the fact that each feature is actually contributing $\cos(0) = 1$ to the total sum, and leads to a distorted picture of the network's predictive strategy.

considers all three channels together; recall here that this rule applies a flat (uniform) rule of relevance distribution in the lowest layers, thus the obtained heatmap is determined by how relevance is distributed spatially across the neurons in the upper layers. The results show that the important features for this prediction form two spatial patterns that are closely aligned with the locations over which the two ARs were detected by the experts (we do not wish to further assess this alignment quantitatively since there is no exact grid-by-grid correspondence between the U850 and the labeled fields). Thus, we can conclude that the network classified this input to the right class based on the wind features that are associated with the labeled ARs locations, which may add to the model's trustworthiness. The XAI results for the other two channels of V850 and integrated precipitable water are presented in Figures S3-S4.

In Figure 8, we consider the same input and use XAI to explain why the CNN assigned a small probability to the class of zero ARs. We again observe that the effect of gradient shattering is drastic and makes the results of Gradient, Smooth Gradient, Input*Gradient, Integrated Gradient and $LRP_z$ incomprehensible. Based on the rest of the methods, and by comparing the Figures 7-8, results show roughly the same patterns but with the opposite sign. This suggests that the features that made the network be certain about the occurrence of two or more ARs are also the features that made the network decide that the considered input is not likely a simulation with zero ARs. Thus, in Figure 8, we verify that the CNN based its decision on features that are associated with the two ARs.

The above results validate the conclusions of our analysis in the previous section and show that the effects observed for the different XAI methods for the synthetic benchmark occur also for climate data and thus need to be taken into account when interpreting the results. In particular, no optimal method exists. Thus, in typical prediction applications, where no ground truth of attribution exists, a more holistic approach should be taken. By considering the explanations from many XAI methods as a whole (as in Figures 7-8) and knowing the relative strengths and weaknesses of each one, scientists may more effectively gain insights about the decision-making strategy of the network, as opposed to the use of a single method.

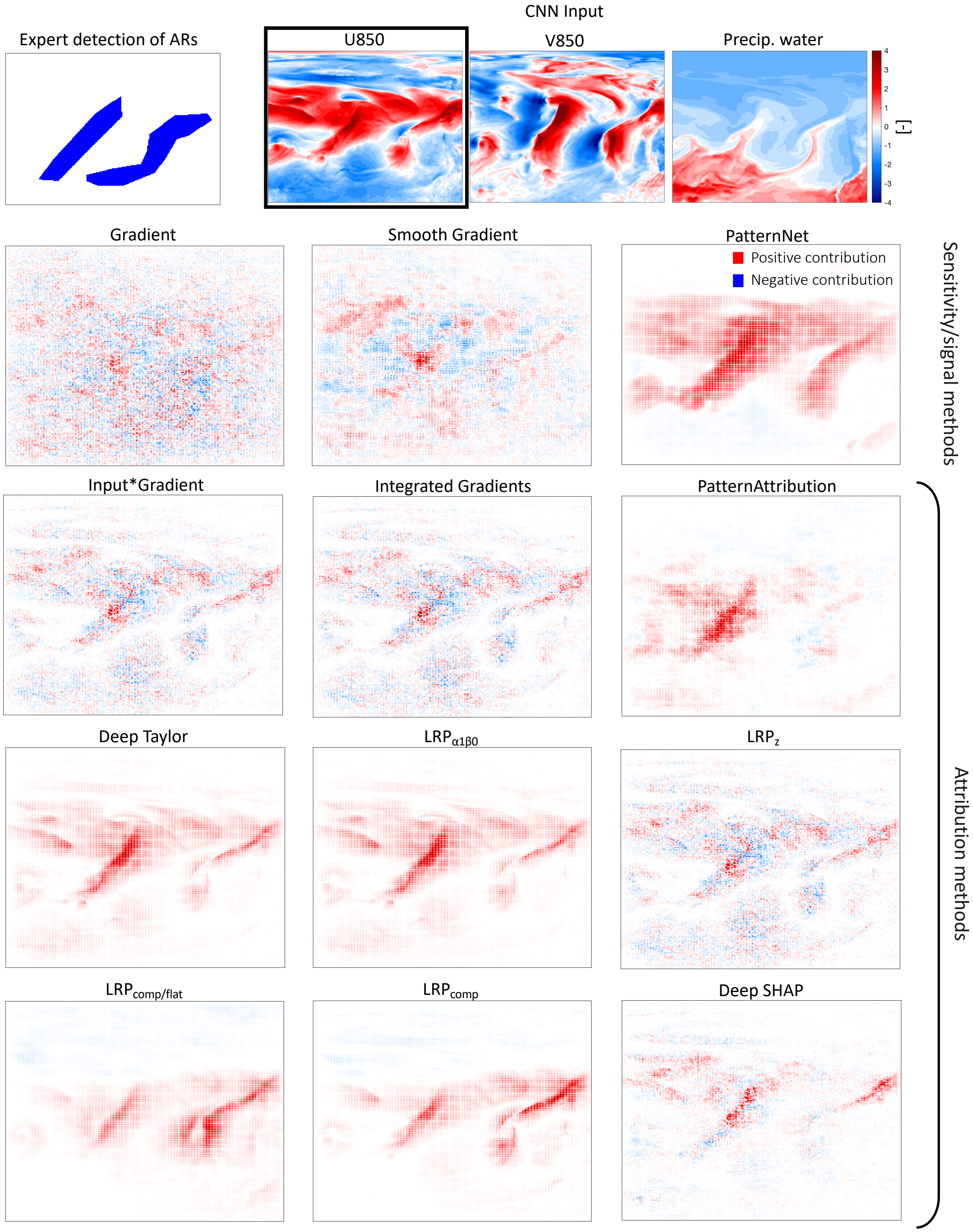

**Figure 7.** Explanations from different XAI methods of the strategy of the CNN for the ClimateNet dataset. The CNN has successfully classified the input image to the class of two or more ARs. XAI methods are applied to explain the successful prediction, and results correspond to the U850 channel. For each heatmap, we divided all values by the maximum (in absolute terms) value.

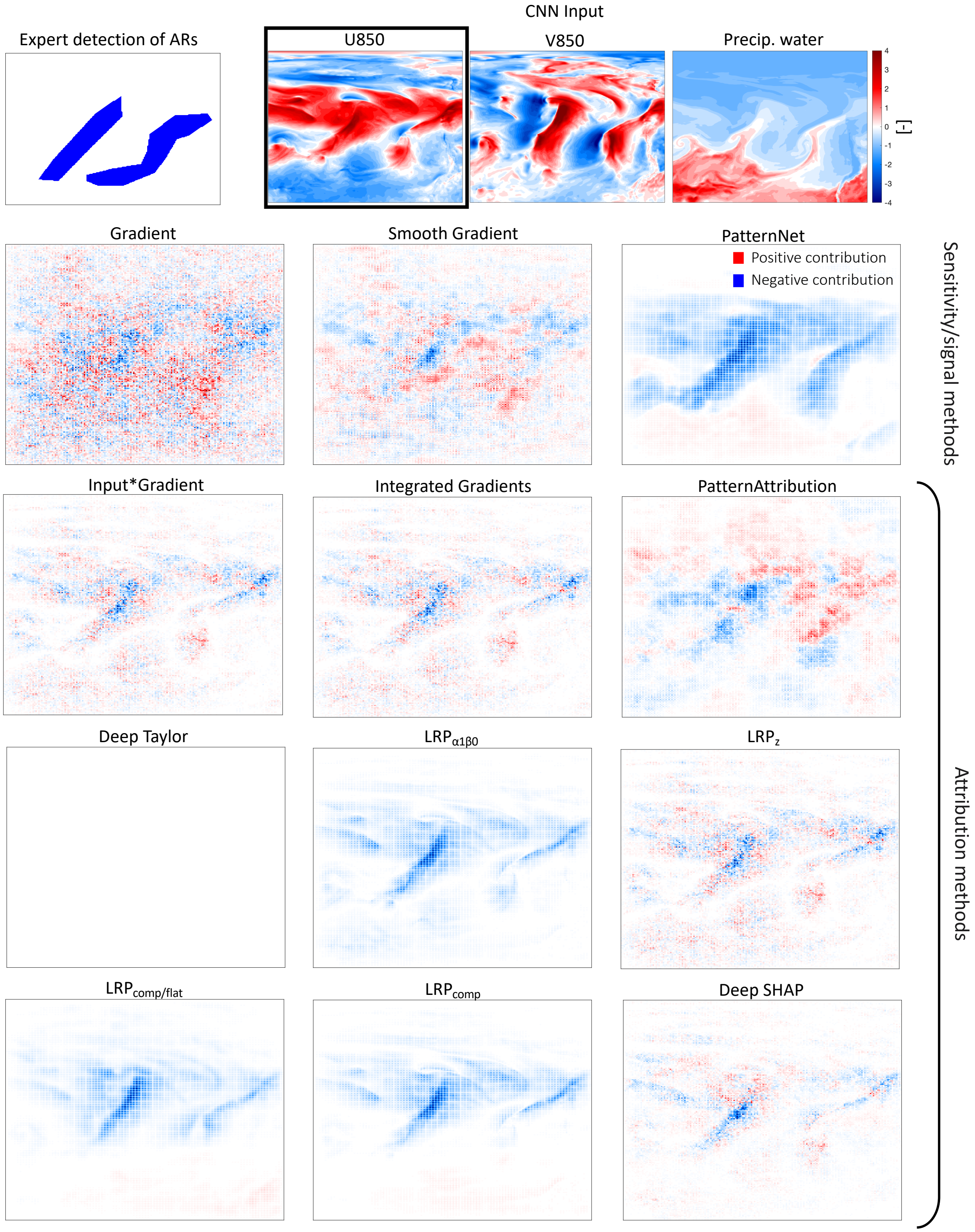

**Figure 8.** Same as in Figure 7, but here XAI methods are applied to explain why the CNN assigned a low probability to the class of zero ARs.

## 4. Conclusions

Explainable artificial intelligence has increasingly been receiving attention in the field of geoscience, as a means to explain black-box models of machine and deep learning that are not inherently interpretable. Although the potential of XAI methods has already been documented in the computer science literature and in geosciences (McGovern et al., 2019; Ebert-Uphoff and Hilburn, 2020; Barnes et al., 2020; Toms et al., 2020; 2021; Sonnewald and Lguensat, 2021; Mayer and Barnes, 2021; Hilburn et al., 2021; Keys et al., 2021; Mamalakis et al., 2022), many studies have highlighted theoretical and practical limitations (Ancona et al., 2018; Kindermans et al., 2017b; Rudin, 2019; Dombrowski et al., 2020; Zhou et al., 2022). Moreover, the assessment of XAI has typically been based on subjective criteria in the recent literature (Mamalakis et al., 2021; Leavitt and Morcos, 2020). To shed more light into the XAI limitations and gain insight into best practices, in this study, we considered some of the most popular XAI methods and compared the fidelity of their explanations in applications for convolutional neural networks relevant to geoscience. To do so, we used a synthetic attribution benchmark, where the ground truth of attribution is *a priori* known, to objectively highlight relative strengths and weaknesses, and a dataset of climate simulations to validate our insights in a more typical prediction setting.

Our investigation revealed aspects that need to be considered when applying XAI methods. These include: i) Gradient shattering (i.e., the phenomenon of noisy patterns in the gradient), the level of which is a function of the depth of the network. For very deep networks, gradient shattering might lead to overwhelmingly noisy patterns that make the explanation of any gradient-based method incomprehensible. ii) Many of the considered methods are either theoretically unable or were shown in practice to be ineffective in disentangling positive and negative contributions. This may lead to a very distorted picture of what the network's strategy is and possibly limit trust in the predictive model itself. iii) Some methods automatically assign a zero attribution to zero values in the input, despite the fact that in specific settings a zero input value could be important for the prediction. We referred to this issue as the "ignorant to zero input" issue. The results of these methods may be more informative if they are viewed as explanations that correspond to a blank image baseline (i.e., an image with only zeros). The effect and/or usefulness of assuming different baselines in XAI research will be the subject of a future study. A summary of the relative strengths and weaknesses that the considered methods exhibit for the types of applications in the current analysis is shown in Table 1.

Our investigation suggests that no optimal method exists for all prediction settings and network architectures. For example, previous studies in computer science and the geosciences has shown that for relatively shallow fully-connected networks and for physical problems where a zero input contains no information, methods like Input*Gradient, Integrated Gradients and $LRP_z$ might perform well (Kohlbrenner et al., 2020; Mamalakis et al., 2021). Yet in this investigation, we showed that for deep CNNs and/or for cases where a zero input might be important for the prediction, these methods might provide a distorted picture of the decision strategy of the network. Having clarified that no universally optimal method exists, we note that for CNN applications, one might have relatively more good reasons to use methods like $LRP_{comp}$, $LRP_{comp/flat}$ and Deep SHAP than others. Yet, these methods are not perfect and require different computational resources, so we would argue that applying many methods and collectively comprehending the CNN strategy (a more holistic approach) is and will be the way to go for the foreseeable future. We conclude by saying that we envision our analysis and revealed insights highlight even more the need for rigorous and objective assessment of XAI methods in order to successfully implement them in geoscience and leverage machine and deep learning for prediction.

## Acknowledgments

This work was supported in part by the National Science Foundation under Grant No. OAC-1934668. I. E.-U. also acknowledges support by the National Science Foundation under Grant No. ICER-2019758. The authors would also like to thank the efforts of the ClimateNet team (see Prabhat et al., 2021) for making their data publicly available.

## Data availability

The code that was used to generate the synthetic data and train and explain the CNN has been made publicly available at: https://github.com/amamalak/XAI_Fidelity_Assessment_CNN_GEO.
The ClimateNet dataset (Prabhat et al., 2021) is publicly available at: https://portal.nersc.gov/project/ClimateNet/.

## Appendix A: The use of additively separable functions for generating synthetic attribution benchmarks and their connection to our study.

As mentioned in section 2.1 of the main text, an attribution benchmark consists of a synthetic input $\mathbf{X}$ and a synthetic output $Y$, with the latter being a known function $F$ of the former (Mamalakis et al., 2021). Regarding the functional form of $F$, this depends on what type of network one wants to benchmark (e.g., a fully connected network, a CNN, etc.), and Mamalakis et al. (2021) noted that the function $F$ can be of any arbitrary choice, as long as it has such a form so that the attribution of any output to the corresponding input is objectively derivable.

Mamalakis et al. (2021) suggested that a simple form for $F$ so that the above property is honored is when $F$ is an *additively separable function*, i.e., there exist local functions $C_i$, with $i = 1, 2, \dots, d$, so that:

$$Y = F(\mathbf{X}) = F(X_1, X_2, \dots, X_d) = C_1(X_1) + C_2(X_2) + \dots + C_d(X_d) \qquad (A.1)$$

where the form of $C_i$ is chosen by the analyst depending on what type of network they want to benchmark. The important think to notice is that because of the summation in Eq. (A.1), and for any form of $C_i$, the contribution of any input feature $X_i$ to the output $y_n$ in the sample $n$ is by definition equal to $C_i(x_{i,n})$; that is when considering a zero baseline. This allows for deriving a ground truth of the attribution for any sample $n$, and for any input feature $X_i$, and thus, a synthetic benchmark with $F$ being an additively separable function allows for objectively benchmarking XAI methods.

As described in section 2.1, in this study, we generated a series of images where circular and square frames are present, and the task was to classify each image depending on which class of frames covers more area. This classification task can be shown to fall under the umbrella of additively separable functions as in Eq. (A.1). Specifically, to generate the synthetic output of the current dataset, we may follow the framework of Mamalakis et al. (2021), and define the output variable $Y \in \mathbb{Z}^*$ as in Eq. (A.1), but where:

$$C_i(x_{i,n}) = \begin{cases} 1, & \text{if } i \text{ belongs to a square frame} \\ -1, & \text{if } i \text{ belongs to a circular frame} \\ 0, & \text{otherwise} \end{cases} \qquad (A.2)$$

By combining Eq. (A.1) and (A.2), one can quickly notice that $Y$ essentially represents the difference of the total area of square frames minus the total area of the circular frames in each image. If $Y > 0$, then the square frames cover more area in the corresponding image, and if $Y < 0$, the circular frames cover more area (note that during the simulation of the synthetic dataset, samples that happen to exhibit $Y = 0$ may be disregarded). Thus, the classification task is simplified to predicting the sign of the output $Y$. A negative sign of $Y$ corresponds to class 1 and a positive sign of $Y$ corresponds to class 2, as these are defined in section 2.1.

The ground truth of the attribution is easily and objectively derivable, following Eq. (A.2). In simple terms, and in accordance to the discussion in section 2.1, for a sample $n$, pixels that belong to any square frames contribute positively to the value of $y_n$ (i.e., these pixels "push" $y_n$ to have a positive sign), while pixels that belong to any circular frames contribute negatively (i.e., these pixels "push" $y_n$ to have a negative sign). We highlight the latter rule of attribution is valid when considering a blank image as the baseline. Moreover, the contribution of each pixel to the output $Y$ depends on whether the pixel belongs to a circular or square frame (see Eq. (A.2)), thus, it depends on the values of the neighboring pixels. This inherent spatial dependency makes a CNN be the most suitable type of network to address this classification task.

## Appendix B: Analytical formulas of the considered XAI methods

**Gradient:** In this method (Simonyan et al., 2014), one calculates the partial derivative of the network's output with respect to each of the input features $X_i$, for the specific sample in question. The relevance (or importance) of the feature at grid point $i$ for the network's prediction of sample $n$, is:

$$R_{i,n} = \left.\frac{\partial \hat{F}}{\partial \mathrm{X}_i}\right|_{\mathrm{X}_i = x_{i,n}} \qquad (B.1)$$

where $\hat{F}$ is the function learned by the CNN, as an approximation to the true function $F$. This method estimates the *sensitivity* of the network's output to the input variable $\mathrm{X}_i$. The motivation for using the Gradient method is that if changing the value, $x_{i,n}$, of a grid point is shown to cause a large change to the CNN output, then that grid point may be relevant for the prediction. Furthermore, calculation of the Gradient is very convenient, as it is readily available in any network training environment, contributing to the method's popularity.

**Smooth Gradient:** This sensitivity method was introduced in (Smilkov et al., 2017) and is very similar to the method Gradient, except that it aims to obtain a more robust estimation of the local derivative by averaging the gradients over a perturbed number of inputs with added noise:

$$R_{i,n} = \frac{1}{m}\sum_{j=1}^{m} \left.\frac{\partial \hat{F}}{\partial \mathrm{X}_i}\right|_{\mathrm{X}_i = x_{i,n} + e_{i,n,j}} \qquad (B.2)$$

where $m$ is the number of perturbations, and $e_{i,n,j}$ comes from a standard Normal Distribution.

**Input*Gradient:** As is evident from its name, this method (Shrikumar et al., 2016; 2017) multiplies the local gradient with the input itself, to get the relevance:

$$R_{i,n} = x_{i,n} * \left.\frac{\partial \hat{F}}{\partial \mathrm{X}_i}\right|_{\mathrm{X}_i = x_{i,n}} \qquad (B.3)$$

This method quantifies the *attribution* of the output to the input. Attribution methods aim to quantify the relative contribution of each input feature to the output value, something that is conceptually different from the sensitivity of the output to the input, as in the previous two methods; for a brief explanation of the difference between attribution and sensitivity see Appendix C.

**Integrated Gradients:** This method (Sundararajan et al., 2017) is also an attribution method similar to Input*Gradient method but aims to account for the fact that in nonlinear problems the derivative is not constant. This method considers a reference (baseline) vector $\hat{\boldsymbol{x}}$, (for which the network's output is zero, i.e., $\hat{F}(\hat{\boldsymbol{x}}) = 0$). Then the relevance is equal to the product of the distance of the input from the reference point with the average of the gradients at points along the straightline path from the reference point to the input:

$$R_{i,n} = \left(x_{i,n} - \hat{x}_i\right) * \frac{1}{m}\sum_{j=1}^{m} \left.\frac{\partial \hat{F}}{\partial X_i}\right|_{X_i = \hat{x}_i + \frac{j}{m}(x_{i,n} - \hat{x}_i)} \qquad (B.4)$$

where $m$ is the number of steps in the Riemann approximation.

**Layer-wise Relevance Propagation (LRP):** LRP (Bach et al., 2015; Samek et al., 2016) is an attribution method that sequentially propagates the prediction $\hat{F}(\mathbf{x}_n)$ (more specifically the before-softmax value) back to neurons of lower layers, obtaining the intermediate relevance for all neurons, until the input layer is reached and the relevance of all input features $R_{i,n}$ is calculated. There are many different rules with which this relevance propagation can be performed. Below we consider the most popular rules for CNNs.

**i) LRP$_z$:** In the LRP$_z$ rule, the back propagation is performed as follows:

$$R_i^{(l)} = \sum_j \frac{z_{ij}}{z_j} R_j^{(l+1)} \qquad (B.5)$$

where $R_j^{(l+1)}$ is the relevance of the neuron $j$ at the upper layer $(l+1)$, and $R_i^{(l)}$ is the relevance of the neuron $i$ at the lower layer $(l)$. The propagation is based on the ratio of the localized preactivations $z_{ij} = w_{ij} x_i$ during prediction time and their respective aggregation $z_j = \sum_i z_{ij} + b_j$ in the neuron $j$. Because this rule might lead to unbounded relevances when $z_j$ approaches zero (Bach et al., 2015), additional advancements have been proposed.
**ii) LRP$_{\alpha\beta}$:** In this rule, positive and negative preactivations $z_{ij}$ are considered separately, so that the denominators are always nonzero:

$$R_i^{(l)} = \sum_j \left( \alpha \frac{{z_{ij}}^+}{{z_j}^+} + \beta \frac{{z_{ij}}^-}{{z_j}^-} \right) R_j^{(l+1)} \qquad (B.6)$$

where

$${z_{ij}}^+ = \begin{cases} z_{ij}; \ z_{ij} > 0 \\ 0 \end{cases} \qquad {z_{ij}}^- = \begin{cases} 0 \\ z_{ij}; \ z_{ij} < 0 \end{cases}$$

In our study, we use the commonly used version of this rule where $\alpha = 1$ and $\beta = 0$, which considers only positive preactivations (Bach et al., 2015).
**iii) LRP$_{comp}$:** Due to the different strengths and weaknesses of the LRP$_z$ and LRP$_{\alpha\beta}$ rules that we discuss in the results section, a composite rule that combines these two rules has been recently suggested in the literature (Kohlbrenner et al., 2020). This composite rule essentially applies the LRP$_z$ rule to propagate the relevance in the fully-connected layers of the CNN, and applies the LRP$_{\alpha\beta}$ rule for the convolutional layers of the CNN. The aim is to combine the strengths and limit the effects of the weaknesses of the two rules. This rule has been suggested as a "best practice" implementation of LRP when explaining a deep CNN (Kohlbrenner et al., 2020).
**iv) LRP$_{comp/flat}$:** This rule is an extension of LRP$_{comp}$. It implements the rules LRP$_z$ and LRP$_{\alpha\beta}$ exactly the same way as the LRP$_{comp}$ rule, but additionally implements a flat rule in the very lowest layer(s). The flat rule distributes the relevance of a neuron uniformly to all connected neurons in the lower layer. It is designed to be used for convolutional layers and it is not suitable for fully connected layers:

$$R_i^{(l)} = \sum_j \frac{1}{\sum_i 1} R_j^{(l+1)} \qquad (B.7)$$

The motivation behind this flat rule is that it allows the user to modify the resolution of the heatmap by changing which layers the flat rule is applied to (e.g., only at the input layer or the lowest three layers, etc.). If the user is not interested in local accuracy, but they only need to obtain a coarse picture of the relevance, this is a suitable rule to use. Another important aspect of this rule is that it is invariant to any transformation of the input (see section 3 and Bach et al., 2016).

**Deep Taylor Decomposition:** For each neuron $j$ at an upper layer $(l+1)$, this attribution method (Montavon et al., 2017) computes a rootpoint $\hat{x}_i^j$ close to the input $x_i$, for which the neuron's relevance is zero, and uses the difference $(x_i - \hat{x}_i^j)$ to estimate the relevance of the lower-layer neurons recursively. The relevance re-distribution is performed as follows:

$$R_i^{(l)} = \sum_j \left. \frac{\partial R_j^{(l+1)}}{\partial x_i} \right|_{x_i = \hat{x}_i^j} * (x_i - \hat{x}_i^j) \qquad (B.8)$$

where $R_j^{(l+1)}$ is the relevance of the neuron $j$ at the upper layer $(l+1)$, and $R_i^{(l)}$ is the relevance of the neuron $i$ at the lower layer $(l)$. It has been shown in (Samek et al., 2016; Montavon et al., 2017) that for neural networks with ReLU activations, Deep Taylor leads to identical results to the $LRP_{\alpha 1 \beta 0}$ rule.

**PatternNet** and **PatternAttribution:** These methods are based on the idea that every input consists of a signal component (all of the information in the input that is relevant to the prediction task) and a distractor (all of the distracting information that is irrelevant to the prediction task). Kindermans et al. (2017a) argued that most existing XAI methods do not necessarily disentangle the signal and the distractor before attributing the output to the input. In fact, the authors showed that even for a simple linear regression model, the vector of weights (i.e., regression coefficients) that is typically used to interpret the model is not necessarily aligned with the direction of the signal in the input (Kindermans et al., 2017a). Thus, Kindermans et al. (2017a) argued that to explain a model, one needs to develop an approach that distinguishes between the signal and the distractor in the input, and they proposed PatternNet to estimate the signal in the input and PatternAttribution to then attribute each prediction to the input. Both methods implement a layer-wise propagation of the prediction back to lower layers until the input layer is reached and the signal or the attribution is obtained (i.e., similar to the LRP method). The propagation rules are:

$$s_i^{(l)} = \sum_j \alpha_{ij}\, s_j^{(l+1)} \qquad (B.9.1)$$

$$R_i^{(l)} = \sum_j w_{ij}\alpha_{ij}\, R_j^{(l+1)} \qquad (B.9.2)$$

for PatternNet and PatternAttribution, respectively. Above, $s_i^{(l)}$ and $R_i^{(l)}$ are the signal and the attribution (relevance) of neuron $i$ in the layer $(l)$. In both methods, the summation over $j$ considers only the neurons in the upper layer $(l+1)$ that were activated in the forward pass of the specific prediction. The symbol $w_{ij}$ represents the weight from neuron $i$ to neuron $j$, while the vector $\boldsymbol{\alpha}_j = \{\alpha_{ij}, \forall i\}^T$ represents the direction of the signal in the neurons of the layer $(l)$ and the neuron $j$ and is estimated using the training dataset as:

$$\boldsymbol{\alpha}_j = \frac{E_+[\mathbf{x}, z_j] - E_+[\mathbf{x}]E[z_j]}{\mathbf{w}_j{}^T E_+[\mathbf{x}, z_j] - \mathbf{w}_j{}^T E_+[\mathbf{x}]E[z_j]} \qquad (B.9.3)$$

where $\mathbf{w}_j = \{w_{ij}, \forall i\}^T$ is the weight vector, $\mathbf{x} = \{x_i, \forall i\}^T$ is the vector with all the activations of the neurons $i$ in the layer $(l)$ and $z_j$ is their linear projection in the neuron $j$. The symbol $E_+$ indicates that the expectation is only taken over those training samples that correspond to positive $z_j$. Note that the expressions in the above ratio represent the covariance of $\mathbf{x}$ and $z_j$.

**Deep SHAP:** Deep SHAP is an attribution method that is based on the use of Shapley values (Shapley, 1953) and is specifically designed for deep neural networks (Lundberg and Lee, 2017). The Shapley values originate from the field of cooperative game theory and represent the average expected marginal contribution of each player in a cooperative game, after all possible combinations of players have been considered (Shapley, 1953). Regarding the importance of Shapley values to XAI, it can be shown (Lundberg and Lee, 2017) that across all *additive feature attribution methods* (a general class of attribution methods that unifies many popular XAI methods like LRP), the only method that satisfies all desired properties of local accuracy, missingness and consistency (see Lundberg and Lee, 2017, for details on these properties) emerges when the feature attributions $\varphi_i$ are equal to the Shapley values:

$$\varphi_i = \sum_{S \subseteq M \setminus \{i\}} \frac{|S|!\,(|M| - |S| - 1)!}{|M|} \left[f_{S\cup\{i\}}\left(x_{S\cup\{i\}}\right) - f_S(x_S)\right] \qquad (B.10)$$

where $M$ is the set of all input features, $M\setminus\{i\}$ is the set $M$, but with the feature $x_i$ being withheld, $|M|$ represents the number of features in $M$, and the expression $f_{S\cup\{i\}}\left(x_{S\cup\{i\}}\right) - f_S(x_S)$ represents the net contribution (effect) of the feature $x_i$ to the outcome of the model $f$, which is calculated as the difference between the model outcome when the feature $x_i$ is present and when it is withheld. Thus, the Shapley value $\varphi_i$ is the (weighted) average contribution of the feature $x_i$ across all possible subsets $S \subseteq M \setminus \{i\}$. Due to computational constraints, Deep SHAP approximates the Shapley values for the entire network by computing the Shapley values for smaller components of the network and propagating them backwards until the input layer is reached (similar in philosophy to LRP, PatternNet and PatternAttribution).

## Appendix C: Sensitivity vs Attribution

When explaining a black-box model to a human, one typically aims to disentangle which input features were important/relevant for a specific prediction made by the model. The way to define what "being an important feature" means is not unique, and different definitions or methods to estimate feature importance can lead to different insights. Two of the most important categories of methods that aim to estimate feature importance are the methods that estimate *sensitivity* and the methods that estimate *attribution*. Here we want to briefly clarify the conceptual difference between the two.

Sensitivity refers to how sensitive the value of the output is to a specific input feature. An obvious way to estimate sensitivity is to calculate the first partial derivative of the network function $\hat{F}$ with respect to the input feature of interest. This is what methods like Gradient and Smooth Gradient aim to do. Attribution, on the other hand, refers to the relative contribution of a specific input feature to the output. When dealing with complex models like deep neural networks, estimating attribution becomes complicated and many methods like the Layer-wise Relevance Propagation, Pattern Attribution, and Deep SHAP have been proposed for this task.

To give an illustrative example of the difference between sensitivity and attribution, let us consider a simple nonlinear function $Y = F(X_1, X_2) = \sin(X_1) + \cos(X_2)$. We can easily calculate that at the point $(X_1, X_2) = (0,0)$, we get $Y_{0,0} = F(0,0) = 0 + 1 = 1$. If we were to explain this output $Y_{0,0}$, i.e., if we were to argue about which feature from $X_1, X_2$ was more important for it, we would get conceptually and numerically different answers using a sensitivity versus an attribution perspective. In terms of sensitivity, the output $Y_{0,0}$ is more sensitive to the value of feature $X_1$, than feature $X_2$, because: $\left.\frac{\partial F}{\partial X_1}\right|_{0,0} = \cos(0) = 1$, while: $\left.\frac{\partial F}{\partial X_2}\right|_{0,0} = -\sin(0) = 0$. In terms of attribution, the opposite is true, i.e., the feature $X_2$ contributes more to prediction $Y_{0,0}$, because: $\sin(X_1)|_{X_1=0} = \sin(0) = 0$, while: $\cos(X_2)|_{X_2=0} = \cos(0) = 1$.

Apart from the numerical difference, the conceptual difference between the sensitivity and attribution can be more clearly realized if we think about the units of the results in the two cases. When estimating sensitivity, the units of the importance or relevance are [units of output/units of input], while when estimating attribution, the units of the results are [units of output]. Thus, these two ways of explaining a black-box model are conceptually (and numerically) different, but they can both be insightful in different ways to a human, thus they are equally valuable.